\documentclass[12pt]{article}
\usepackage{graphicx}
\newcommand{\anti}[1]
{
	\overline{#1}\mbox{}
}

\newcommand\pubnumber{SNSN-323-63}
\newcommand\pubdate{\today}

\def\krakow{Institute of Nuclear Physics Krakow\\
ul. Radzikowskiego 152, PL-30 156 Krak\'ow, Poland}

\def\Title#1{\begin{center} {\Large #1 } \end{center}}
\def\Author#1{\begin{center}{ \sc #1} \end{center}}
\def\Address#1{\begin{center}{ \it #1} \end{center}}

\newcommand\pubblock{\rightline{\begin{tabular}{l} \pubnumber\\
         \pubdate  \end{tabular}}}
\newenvironment{Abstract}{\begin{quotation}  }{\end{quotation}}
\newenvironment{Presented}{\begin{quotation} \begin{center} 
             PRESENTED AT\end{center}\bigskip 
      \begin{center}\begin{large}}{\end{large}\end{center} \end{quotation}}
\def\Acknowledgements{\bigskip  \bigskip \begin{center} \begin{large}
             \bf ACKNOWLEDGEMENTS \end{large}\end{center}}

\def\beq{\begin{equation}}
\def\eeq#1{\label{#1}\end{equation}}
\def\eeqn{\end{equation}}

\def\beqa{\begin{eqnarray}}
\def\eeqa#1{\label{#1}\end{eqnarray}}
\def\eeqan{\end{eqnarray}}

\let\bar=\overbar

\def\Dslash{\not{\hbox{\kern-4pt $D$}}}
\def\dslash{\not{\hbox{\kern-2pt $\del$}}}

\def\msb{{\bar{\ssstyle M \kern -1pt S}}}

\begin{document}
\begin{titlepage}
\pubblock

\vfill
\Title{The experimental review of $B \to {D}^{(*)} \tau \nu_{\tau}$ decays.}
\vfill
\Author{ A.Bozek}
\Address{\krakow}
\vfill
\begin{Abstract}
Experimental studies of  $B \to {D}^{(*)} \tau \nu_{\tau}$ decays, are reported. 
The results are based on large data samples collected at the $\Upsilon (4S)$~resonance 
with the Belle detector at KEKB and the BABAR detector at the SLAC PEP-II
  asymmetric energy $e^+ e^-$~colliders.
\end{Abstract}
\vfill
\begin{Presented}
The 6th International Workshop on the CKM Unitarity Triangle, 
University of Warwick, UK, 6-10 September 2010
\end{Presented}
\vfill
\end{titlepage}
\setcounter{footnote}{0}

\section{Introduction}
$B$ decays to $\tau$ leptons represent a broad class of processes that 
can provide interesting 
tests of the Standard Model (SM) and its extensions. Of particular interest 
are the modes presented in this report, the semi-leptonic decays $B \to \bar{D}^{(*)} \tau^+ \nu_{\tau}$\cite{pub-dtaunu,pub-dtaunu2,babar-dtaunu,conf-paper}\footnote{Charge conjugate modes are implied throughout this report unless 
otherwise stated.}. 

In the SM semileptonic 
$B$ decays to $\tau$ leptons occur at tree level.
Branching fractions
are predicted to be smaller than those to light leptons.
The  predicted branching fractions, based on the SM,
are around 1.4\% and 0.7\% 
for $B^0 \to {D}^{*-} \tau ^+ \nu_{\tau}$ 
and $B^0 \to {D}^- \tau ^+ \nu_{\tau}$, 
respectively (see {\it e.g.}, \cite{hwang}).
 
$ B$ meson decays with $b \to  c \tau \nu_{\tau}$ transitions,
due to the large mass of the $\tau$ lepton, are sensitive
probes of models with extended Higgs sectors\cite{taunu-th}\cite{Itoh}.
The semileptonic $B$ decays to tau provide new observables sensitive 
to New Physics such as polarizations, which cannot be accessed in 
leptonic $B$ decays. 
In multi-Higgs doublet models, substantial departures from the
SM decay rate could occur for $B \to \bar{D} \tau^+ \nu_{\tau}$.
Smaller departures are expected for $B \to \bar{D}^{*} \tau ^+ \nu_{\tau}$,
however they provide cleaner sample
and $D^*$ polarisation that can be used to enhance a sensitivity to 
NP effects.

Difficulties 
related to multiple neutrinos in the final states cause that there is 
little experimental 
information about decays of this type. Prior to the B-factories era,
 there was only inclusive measurement of 
$\mathcal{B}(B \to c \tau^+ \nu_{\tau}) = (2.48 \pm 0.26 )\%$ from LEP\cite{PDG2006}.

\section{Analysis techniques}

At $B$-factories $B$ decays to multi-neutrino final states can be observed via 
the recoil of accompanying $B$ meson ($B_{\rm tag}$). 
The $B_{\rm tag}$ can be reconstructed inclusively from all the particles 
that remain after selecting
$B_{\rm sig}$ candidates or exclusively in several hadronic decay modes.
The remaining charged particles
and photons are required to be consistent with the hypothesis that they are 
coming from $B \to D^{(*)} \tau \nu_{\tau}$ decays.
Choice of the $\tau$, $D$ or $D^{*}$ decay modes, as well as the methods of the $B_{\rm tag}$ reconstruction, depend on 
particular analysis requirements on purity and signal extraction procedure, etc.


\subsection{Exclusive reconstruction of $B_{\rm tag}$ in hadronic modes}

In  BaBar results the $B_{\rm tag}$ candidates are reconstructed 
in $1114$ final states $B_{\rm tag} \to D^{(*)} Y^{\pm}$. 
These final states arise from the large number of ways
to reconstruct the $D$ and $D^*$ mesons within the $B_{\rm tag}$
candidate and the possible pion and kaon combinations
within the $Y^±$ system. 
The $Y^{\pm}$ system may consist of
up to six light hadrons ($\pi^{\pm}, \pi^0 , K^\pm$, or $K_S$ ). 

For Belle case, the  $B_{\rm tag}$ candidates are reconstructed in the following 
decay modes: $B^{+}_{\rm tag} \to \overline{D}{}^{(*)0} h^{+}$, and 
$B^{0}_{\rm tag} \rightarrow \overline{D}{}^{(*)-} h^{+}$, where $h^{+}$ can be $\pi^{+},\rho^{+},a_{1}^{+}$ 
or $D_{s}^{(*)+}$. 

The selection of $B_{\rm tag}$ candidates is based on the energy substituted mass
$m_{\rm ES}\equiv\sqrt{E_{\rm beam}^{2} - p_{B}^{2}}$ (called $m_{\rm bc}$ in Belle)
and the energy difference $\Delta E\equiv E_{B} - E_{\rm beam}$.
Here, $E_{B}$ and $p_{B}$ are the reconstructed energy and momentum
of the $B_{\rm tag}$ candidate in the $e^+e^-$ center-of-mass (CM) system,
and $E_{\rm beam}$ is the beam energy in the CM frame.

\subsection{Inclusive reconstruction of $B_{\rm tag}$ in hadronic modes}
\label{inclusive-tag}

The inclusive tagging  was, up to now, exploited only by 
Belle collaboration\cite{pub-dtaunu,pub-dtaunu2}.

With this method the reconstruction starts from $B_{\rm sig}$ candidates.
Reconstruction of $D^{(*)}$ on the signal side 
 strongly suppresses 
the combinatorial and continuum backgrounds.
Once a $B_{\rm sig}$ candidate is found, the remaining particles that 
are not assigned
to $B_{\rm sig}$
are 
used to reconstruct the $B_{\rm tag}$ decay.  The consistency of a 
$B_{\rm tag}$ 
candidate 
with a $B$-meson decay is checked using the beam-energy constrained mass and 
the energy difference variables:
$M_{\rm tag} = \sqrt{E^2_{\rm beam} - {\bf p}^2_{\rm tag}}$, 
${\bf p}_{\rm tag} = \sum_i {\bf p}_i$,
and
$\Delta E_{\rm tag} = E_{\rm tag} - E_{\rm beam}, ~~ E_{\rm tag} 
= \sum_i E_i$,
${\bf p}_i$ and $E_i$ 
denote the 3-momentum vector and energy of the $i$'th particle.
All quantities are evaluated in the $\Upsilon(4S)$ rest frame.
The summation is over all particles that are left after reconstruction of 
$B_{\rm sig}$ candidates.

To suppress background and improve 
the quality of the $B_{\rm tag}$ selection, additional requirements are imposed
 like:
zero total event charge;     
no charged leptons in tag side;
zero net proton/anti-proton number. 
The requirement  of the high missing mass 
 results in flat $M_{\rm tag}$ distributions for 
most background components , 
while the distribution of the signal modes
peaks, at the $B$ mass.
The main sources of the peaking 
background are
the semileptonic decays 
$B \to \bar{D}^{*}l^+\nu_l$
and $B \to \bar{D}^{(*)}\pi l^+\nu_l$
(including $\bar{D}^{**}l^+\nu_l$).

\section{$B \to D^{(*)} \tau^+ \nu_{\tau}$ with inclusive hadronic tag}
 Belle collaboration reported the first observation 
of an exclusive decay with the $b \to c \tau \nu_{\tau}$ transition\cite{pub-dtaunu}, in the 
$B^0 \to D^{*+} \tau^- \nu_{\tau}$ channel using inclusive  $B_{\rm tag}$ 
reconstruction in a data sample containing $535 \times 10^6$ $B\bar{B}$ pairs.
The $\tau^- \to e^- \nu_e \nu_{\tau}$ and $\tau^- \to \pi^- \nu_{\tau}$ modes are 
used to reconstruct $\tau$ lepton candidates. 

The observed signal of  $60^{+12}_{-11}$  events for the decay 
$B^0\to D^{*-}\tau^+\nu_{\tau}$
was extracted from $M_{\rm tag}$ distribution.

A new analysis for $B^+ \to D^{(*)0} \tau^+ \nu_{\tau}$ was performed in  a sample of  
657$\times 10^6~B\bar{B}$ pairs\cite{pub-dtaunu2}.
The signal and combinatorial background yields are extracted from an extended
unbinned maximum likelihood fit to the
$M_{\rm tag}$ and $P_{D^0}$ (momentum of $D^0$ from $B_{\rm sig}$
measured in the $\Upsilon(4S)$ frame) variables.  
The $\tau^+ \to e^+ \nu_e \nu_{\tau}$, $\tau^+ \to \pi^+ \nu_{\tau}$, and 
in addition $\tau^+ \to \mu^+ \nu_{\tau}$ modes are used to reconstruct $\tau$ 
lepton candidates.
In total, 13 different decay chains are considered, eight with $\bar{D}^{*0}$
and five with $\bar{D}^0$ in the final states.
The fits are
performed 
simultaneously to all data subsets.
In each of the sub-channels, the data was described as the sum of
four components: signal, cross-feed between  
$\bar{D}^{*0}\tau^+\nu_{\tau}$ and $\bar{D}^{0}\tau^+\nu_{\tau}$, 
combinatorial and peaking 
backgrounds.
The common signal branching fractions
$\mathcal{B}(B^+\to \bar{D}^{*0}\tau^+\nu_{\tau})$
and $\mathcal{B}(B^+\to \bar{D}^{0}\tau^+\nu_{\tau})$,
and the numbers of combinatorial background in each sub-channel 
are free parameters of the fit,
while the normalisations of peaking background contributions are 
fixed 
to the values 
obtained 
from the rescaled MC samples. 
The signal yields and branching fractions for 
$B^+\to \bar{D}^{(*)0}\tau^+\nu_{\tau}$ decays are
related  assuming 
equal fractions of charged and 
neutral $B$ meson pairs produced in $\Upsilon(4S)$ decays.
All the intermediate branching fractions are taken from the 
PDG compilation \cite{PDG2006}. 

The signal yields are 
$446^{+58}_{-56}$ 
$B^+\to \bar{D}^{*0} \tau^+ \nu_{\tau}$ events
and $146^{+42}_{-41}$ 
$B^+\to \bar{D}^{0} \tau^+ \nu_{\tau}$ events.
\begin{figure}[htb]
\centering
\includegraphics[height=1.5in,width=0.24\textwidth]{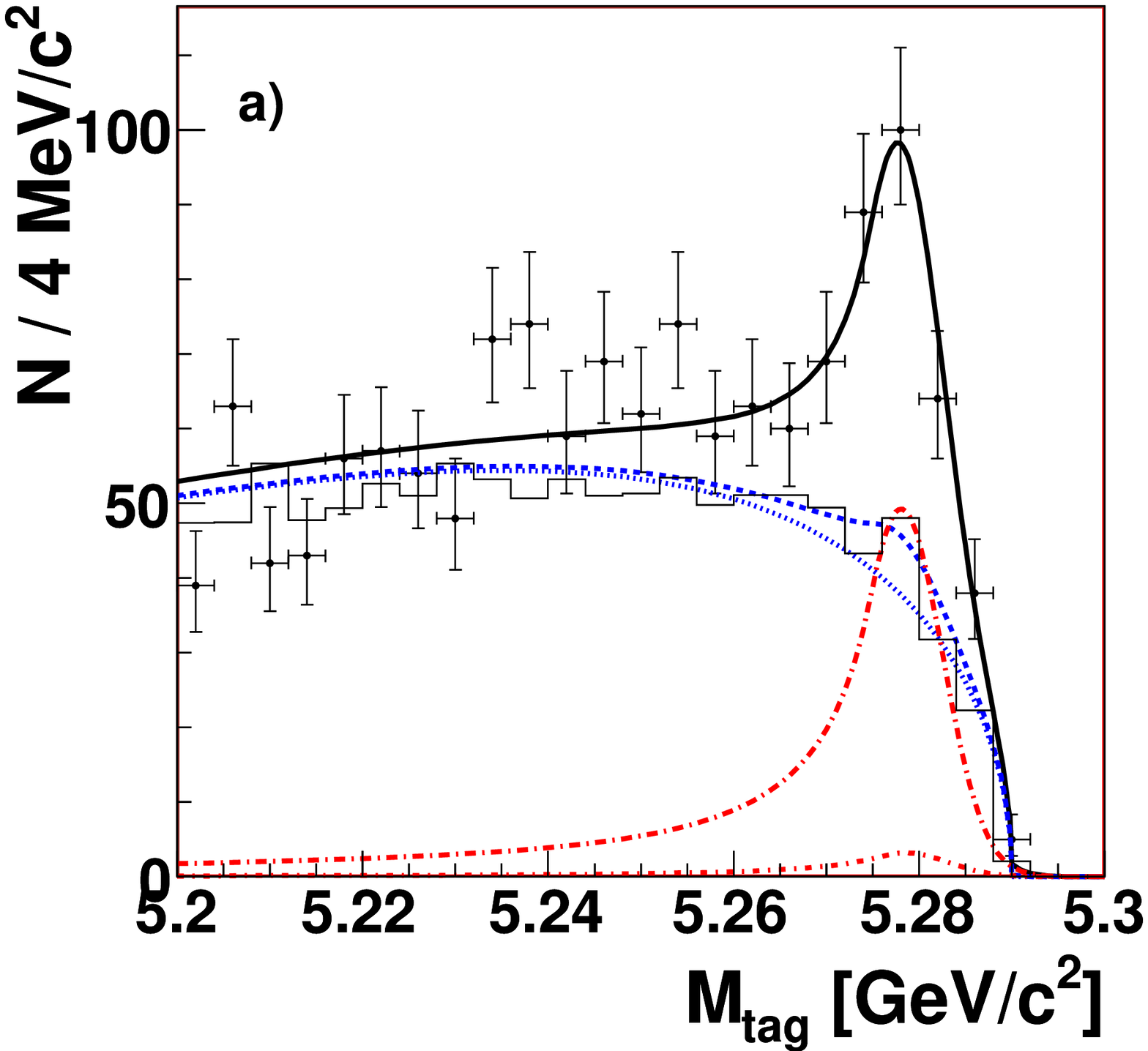}
\includegraphics[height=1.5in,width=0.24\textwidth]{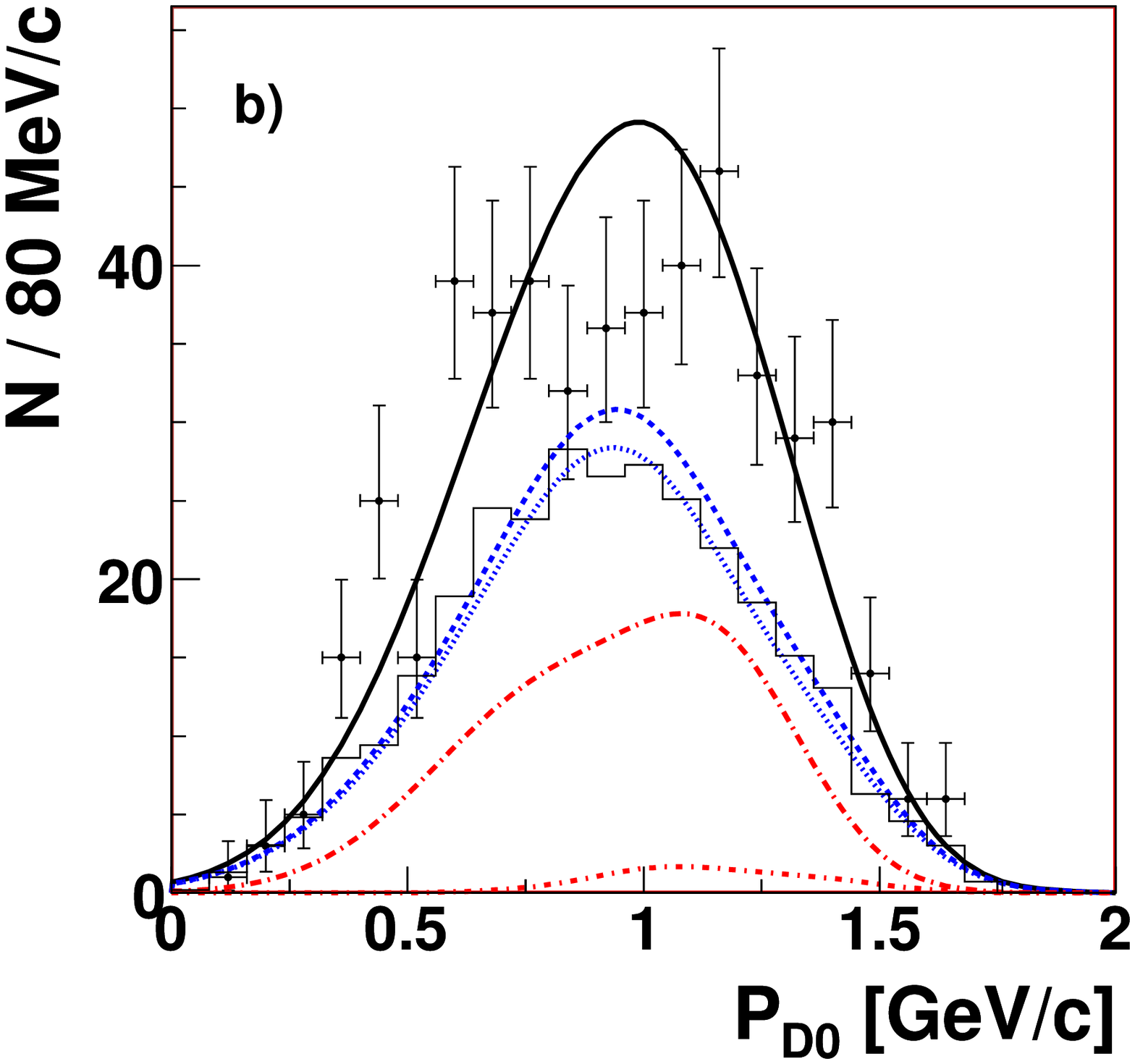}
\includegraphics[height=1.5in,width=0.24\textwidth]{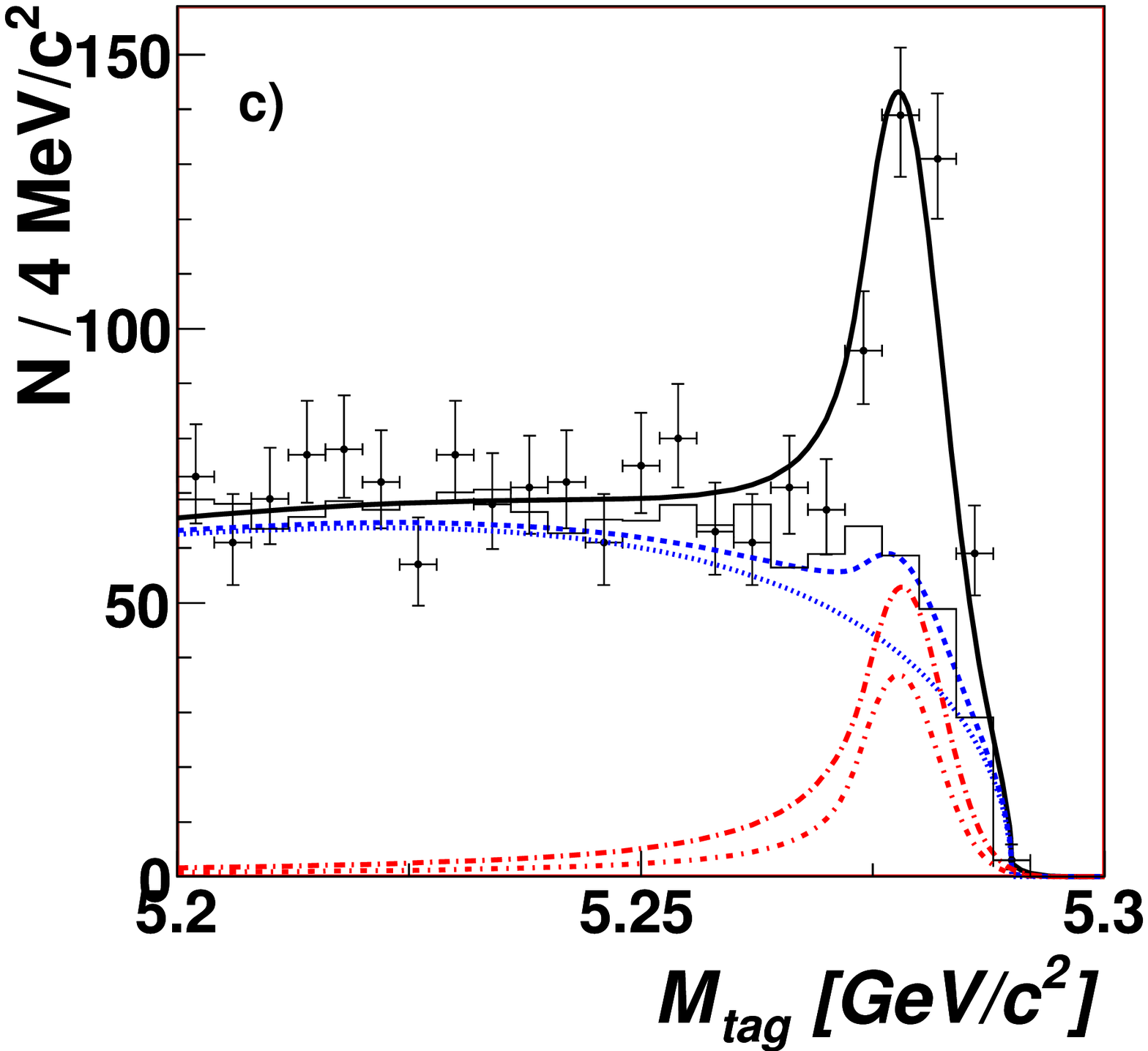}
\includegraphics[height=1.5in,width=0.24\textwidth]{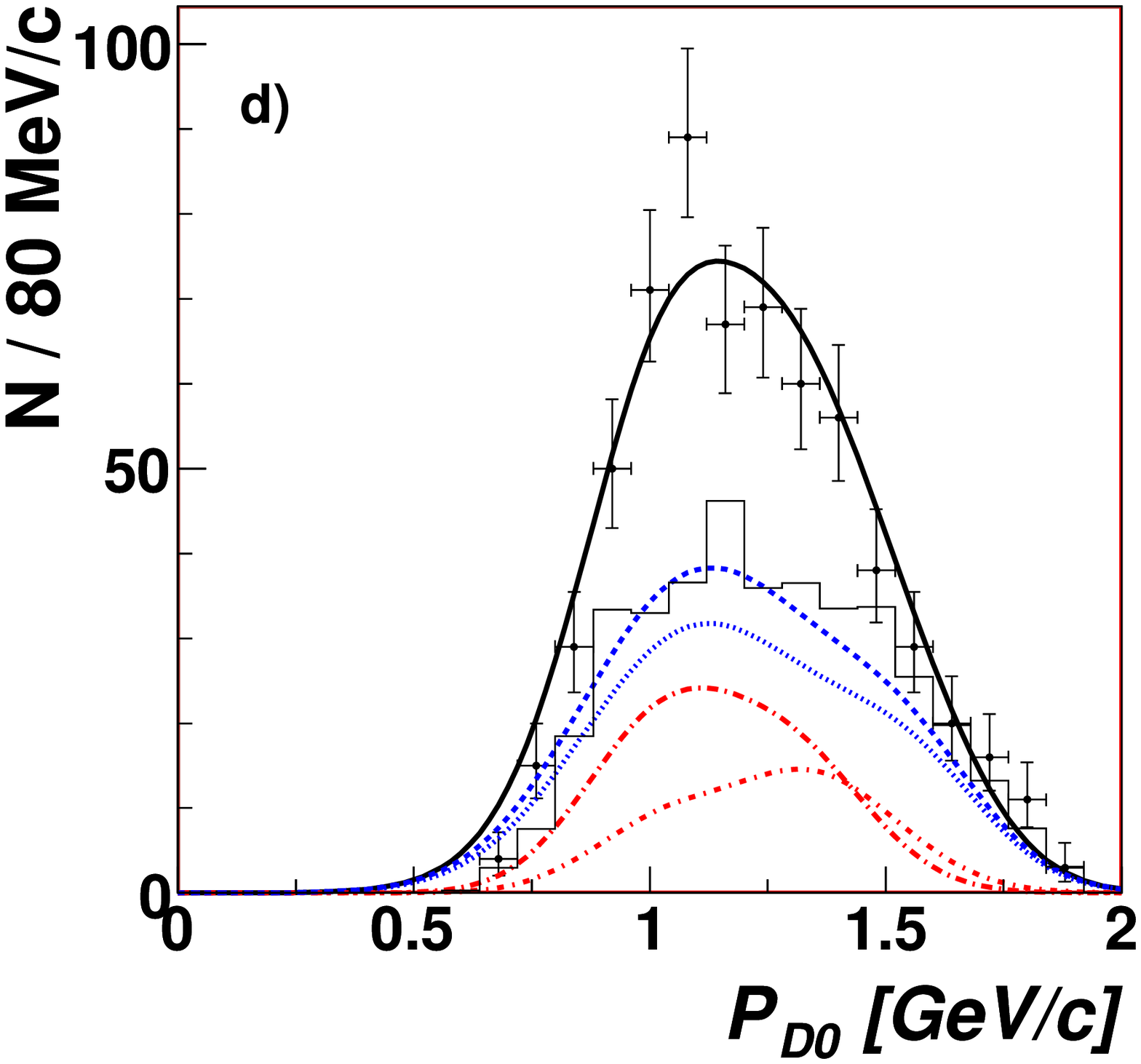}
\caption{The fit projections to
$M_{\rm tag}$,
and $P_{D^0}$ for $M_{\rm tag}>5.26 ~{\rm GeV}/c^2$  
(a,b) for $\bar{D}^{*0}\tau^+\nu_{\tau}$, (c,d)
for $\bar{D}^{0}\tau^+\nu_{\tau}$.}
\label{fig:fit}
\end{figure}


\section{$B \to D^{(*)} \tau^+ \nu_{\tau}$ with exclusive hadronic tags}

Babar collaboration presented  measurements of the semileptonic decays 
$B^- \to D^{0} \tau^- \bar{\nu_{\tau}}$, $B^- \to D^{*0} \tau^- \bar{\nu_{\tau}}$,
 $B^0 \to D^{+} \tau^- \bar{\nu_{\tau}}$, $B^0 \to D^{+} \tau^- \bar{\nu_{\tau}}$,
 and  $B^0 \to D^{*+} \tau^- \bar{\nu_{\tau}}$\cite{babar-dtaunu}.
The data sample consists of $232 \times 10^6$ $\Upsilon(4S) \to B\bar{B}$ 
decays.
The events are selected with a $D$ or $D^{*}$ meson and a light lepton
($ = e$ or $\mu$) recoiling against a fully reconstructed $B$ meson.

\begin{center}
\begin{figure}[]
\vspace{-1.5pc}
\begin{center}
\begin{minipage}{10pc}
\begin{center}
\includegraphics[width=10pc,keepaspectratio,clip]{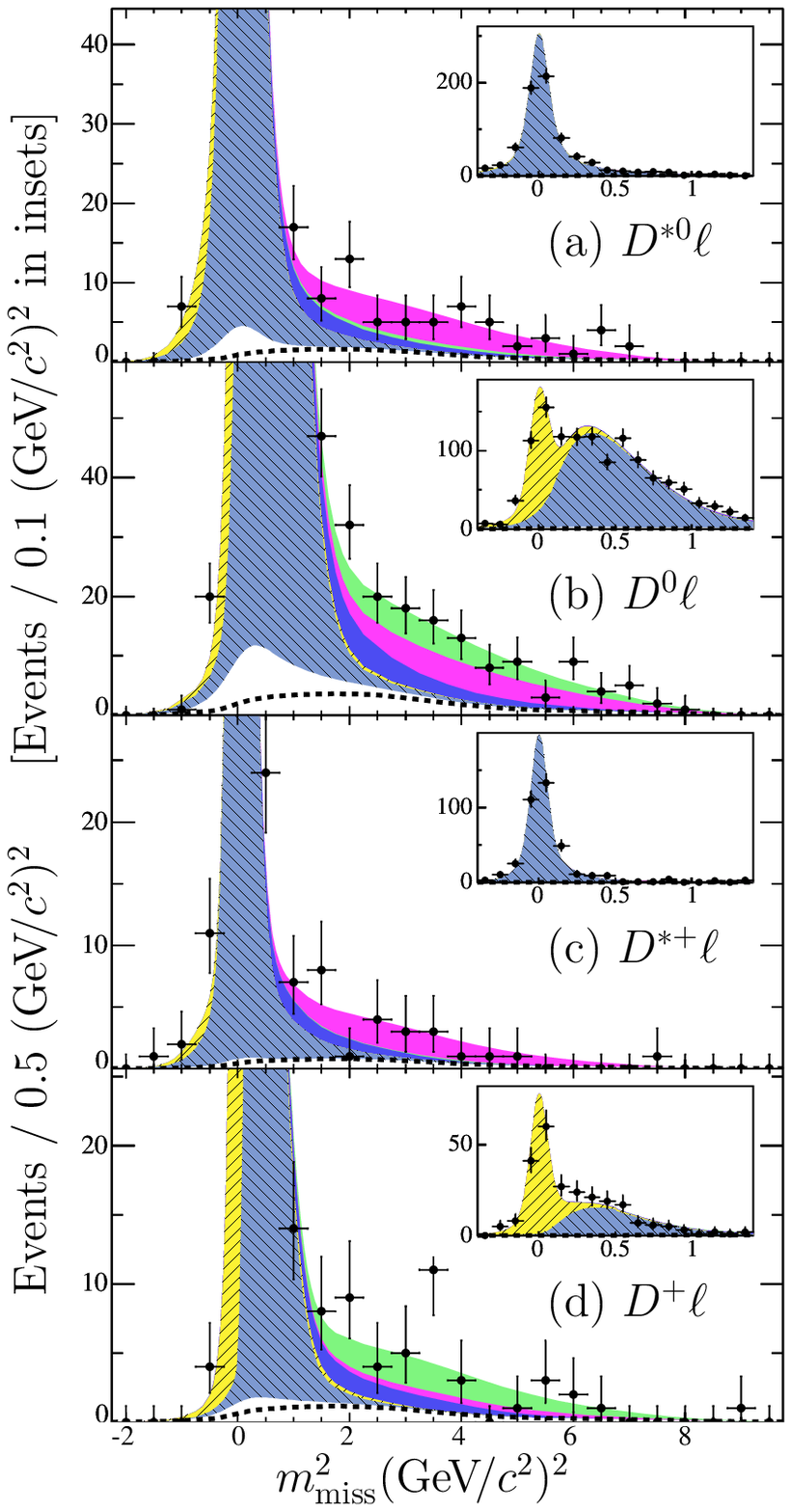}
\end{center}
\end{minipage}
\vspace{-3.0pc}
\hspace{1.5pc}%
\begin{minipage}{10pc}
\begin{center}
\includegraphics[width=10pc,keepaspectratio,clip]{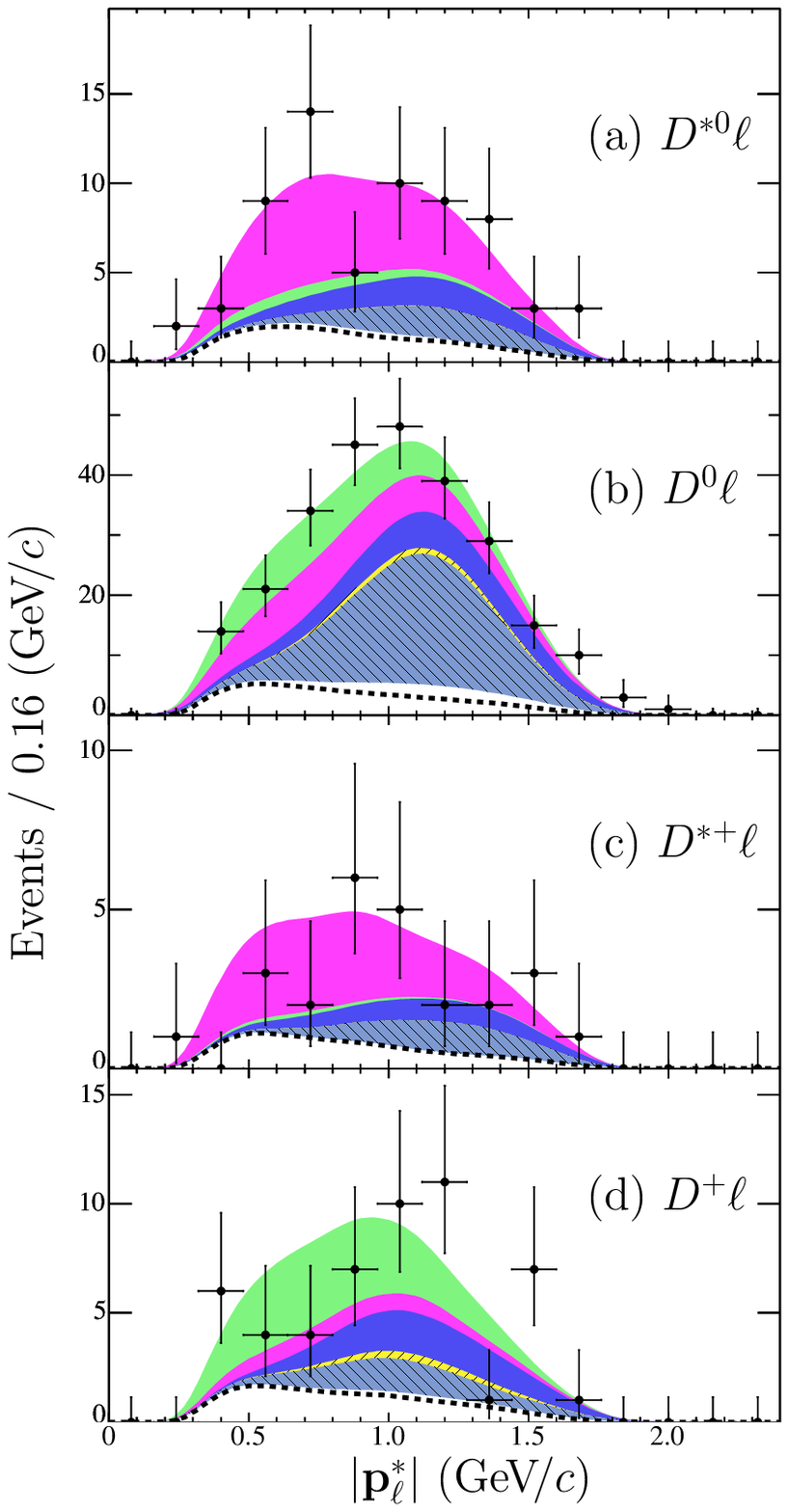}
\end{center}
\end{minipage} 
\vspace{2.0pc}
\caption{\label{pic-babar-dtaunu}\label{pic-babar-dtaunu1} Distributions of events and fit 
projections in $m^2_{\rm miss}$(left) and in $|P^*_l|$ for the four final states: $D^{*0}l^-$, $D^0l^-$, 
$D^{*+}l^-$, and $D^+l^-$. The normalisation region $m^2_{\rm miss} \approx 0$ is 
shown with finer binning in the insets. The $|P^*_l|$ is shown in the signal region, $ m^2_{\rm miss}>1 (GeV/c^2)^2$.
The fit components are combinatorial 
background (white, below dashed line), charge-crossfeed
background (white, above dashed line), the $B \to D l^− \nu_l$ 
normalisation mode (// hatching, yellow), the $B \to D^* l^− \nu_l$ 
normalisation mode (\textbackslash\textbackslash  hatching, light blue), $B \to D^{**} l^− \nu_l$
background (dark, or blue), the $B \to D \tau^− \nu_{\tau}$ signal (light
grey, green), and the $B \to D^* \tau^− \nu_{\tau}$ signal 
(medium grey, magenta).
}
\end{center}
\end{figure}
\end{center}

The fit is performed  to the joint distribution of lepton momentum and 
missing mass squared,~$m_{\rm miss}^2$, to distinguish signal $B \to D^{(*)} \tau^- \nu_{\tau}
(\tau^- \to l^- \bar{\nu_l} \nu_{\tau})$
events from the backgrounds, predominantly $B \to D^{(*)} \ell^- \bar{\nu_{\ell}}$.
The fit is performed simultaneously in four signal channels.
Figure~\ref{pic-babar-dtaunu} shows projections in $m^2_{\rm miss}$
for the four signal channels, showing both the low $m^2_{\rm miss}$
region, which is dominated by the normalisation modes
$B \to D^{(*)} l^− \nu_l$ , and the high $m^2_{\rm miss}$ region, which is 
dominated by the signal mode $B \to D^{(*)} \tau^− \nu_{\tau}$. 

Babar measures the branching-fraction
ratios $R(D) = B(B \to  D \tau^− \nu_{\tau} )/B(B \to D l^− \nu_l )$ and 
$R(D^* ) = B(B \to  D^* \tau^− \nu_{\tau} )/B(B \to D^* l^− \nu_l )$
and, from a combined fit to $B^+$ and $B^0$ channels, approximately
$67$ $B \to  D \tau^− \nu_{\tau}$  and $101$ $B \to  D^* \tau^− \nu_{\tau}$ signal events are observed,
corresponding to resulting ratios 
$R(D) = (41.6 \pm 11.7 \pm 5.2) \%$ and
$R(D^*) = (29.7 \pm 5.6 \pm 1.8) \%$, where the uncertainties are statistical 
and systematic. The signal significances are $3.6\sigma$ and $6.2\sigma$ for 
$R(D)$ and $R(D^*)$, respectively.  Normalising to world averaged $B^- \to D^{(*)0} l^- \nu_l$ branching 
fractions\cite{PDG2006}, they obtain $B(B \to  D^* \tau^− \nu_{\tau} ) = (0.86 \pm 0.24 \pm
0.11 \pm 0.06)\%$ and $B(B \to  D^* \tau^− \nu_{\tau}) = (1.62 \pm 0.31 \pm 0.10 \pm 0.05)\%$, where the additional third
uncertainty is from the normalisation mode. They present for the first time, 
distributions of the lepton momentum, $|p^*_l |$, and the squared momentum 
transfer, $q^2$.

Belle collaboration presented similar study based on  604.5 $fb^{−1}$ of
the data sample\cite{conf-paper}.
The $B \to D \tau \nu_{\tau}$ and $B \to D^{*} \tau \nu_{\tau}$ signals are extracted
using unbinned extended maximum likelihood fits to the two-dimensional
$(m_{\rm miss}^2, E_{\rm extra}^{\rm ECL})$ distributions obtained after the 
selection of the signal decays. The $B^+$ and $B^0$ samples are fitted
separetly.
The cross talk between the two tags is found to be small. 
Then for each $B^0$ and $B^+$ tag, a fit is performed simultaneously to 
the two distributions for the $D \tau \nu_{\tau}$ and $D^* \tau \nu_{\tau}$.
The fit components are two signal modes; $B \to D \tau \nu_{\tau}$ and 
$B \to D^{*} \tau \nu_{\tau}$, and the backgrounds from $B \to D \ell \nu_{e}$,
$B \to D^* \ell \nu_{e}$ and other processes.
For the fitting of the $B^0 \to D^{*-} \tau^+ \nu_{\tau}$ distribution, 
the $D \tau \nu_{\tau}$ cross feed  and $D \ell \nu_{\ell}$ background 
 are not included, because their contribution are found 
to be small.


The results 
for the four ratios are listed in Table\ref{table-1}, 
Taking into account the branching fractions for the $B \to D^{*} \ell \nu_{l}$ normalisation 
decays, reported in \cite{PDG2006}
, the branching fractions for the $B \to D^{*} \tau \nu_{\tau}$ decays are obtained and listed in Table\ref{table-2}.
\subsection{Summary} 

Experimentally all modes are clearly established, with significance at 
least $3 \sigma$ (over $5 \sigma$ for $D^{*}$ modes). They are observed in both 
experiments and there is still a room for improvement since the results are not 
based on full statistics.

\begin{center}
\begin{table*}[hbt]
\caption{\label{table-2}Summary of  branching-fractions for $B\to D^{(*)} \tau \nu_{\tau}$ decays($[\%]$), where the first error is statistical, the second is 
systematic, and the third is due to the branching fraction error for 
the normalisation modes. In brackets are significances, after including the systematics$([\sigma])$. }
\begin{tabular}{|l|c|c|c|}
\hline
Mode   & Belle\cite{pub-dtaunu}\cite{pub-dtaunu2}& BaBar\cite{babar-dtaunu}  & Belle\cite{conf-paper}  \\
\hline
$\anti{D}^{*0}$ &$2.12^{+0.28}_{-0.27} 
 \pm 0.29(8.1)$ & $2.25\pm0.48\pm0.22\pm0.17(5.3)$ & $3.04 ~^{+0.69}_{-0.66}~^{+0.40}_{-0.47} \pm 0.22(3.9)$\\
$D^{*-}$& $2.02^{+0.40}_{-0.37} \pm 0.37(5.2)$&$1.11\pm0.51\pm0.04\pm0.04(2.7) $& $2.56~^{+0.75}_{-0.66} ~^{+0.31}_{-0.22} \pm 0.10(4.7)$ \\
$\anti{D}^0$& $0.77 \pm 0.22 
 \pm 0.12(3.5) $& $0.67\pm0.37\pm0.11\pm0.07(1.8)$& $1.51 ~^{+0.41}_{-0.39}~^{+0.24}_{-0.19} \pm 0.15(3.8)$\\
$D^-$&-&$1.04\pm0.35\pm0.15\pm0.10(3.3)$ &$1.01 ~^{+0.46}_{-0.41}  ~^{+0.13}_{-0.11} \pm 0.10(2.6)$ \\
\hline
\end{tabular}
\end{table*}
\end{center}
\begin{center}
\begin{table*}[hbt]
\caption{\label{table-1}The measured  branching-fraction ratios for 
individual $D^{(*)}$ states for analysis based on exclusive $B_{\rm tag}$ reconstruction.The first errors are the statistical and the second errors are the
systematic.
}
\begin{center}
\begin{tabular}{|l|c|c|}
\hline
   & BaBar\cite{babar-dtaunu} & Belle\cite{conf-paper} \\
\hline
$R(\overline{D}{}^0) $ & $(31.4 \pm 17.0 \pm 4.9 )\%$ & $(70~^{+19}_{-18}~^{+11}_{-9})\%$\\  
\hline
$R(\overline{D}{}^{*0}) $ & $(34.6 \pm 7.3 \pm 3.4 )\%$ & $ (47~^{+11}_{-10}~^{+6}_{-7}) \%$\\
\hline
$R(D^-)$ & $(48.9 \pm 16.5 \pm 6.9 )\%$ & $(48~^{+22}_{-19}~^{+6}_{-5}) \%$ \\
\hline
$R(D^{*-})$ & $(20.7 \pm 9.5 \pm 0.8)\%$ & $(48~^{+14}_{-12}~^{+6}_{-4}) \%$\\
\hline
\end{tabular}
\end{center}
\label{tab-bf}
\end{table*}
\end{center}

The current experimental status of semi tauonic B decays is summarized in Tables~\ref{table-1} and \ref{table-2}. 

There is no yet HFAG experimental average of the semi-tauonic 
$B$ decays. Taking into account all available experimental results 
from Belle and Babar a  naive weighted averages can be calculated\footnote{it takes into account correlations in systematic for Belle results}:
	\begin{itemize}
		\item	$\mathcal{B} ( B^+ \to \anti{D}^{*0} \tau^+ \nu_\tau  )=(2.36 \pm 0.27)\% $
	\item	$\mathcal{B} ( B^0 \to D^{*-} \tau^+ \nu_\tau  )=(1.70 \pm 0.34)\% $ 
	\item	$\mathcal{B} ( B^+ \to \anti{D}^0 \tau^+ \nu_\tau )=(0.89 \pm 0.20) \%$    	
	\item	$\mathcal{B} ( B^0 \to D^- \tau^+ \nu_\tau )=(1.03 \pm 0.30)\%$	
	\end{itemize}

These results are consistent with the SM but, given the uncertainties, there is still a room
for a sizeable non-SM contribution. The Super B-factories with $\approx 50$ times higher statistics should measure these modes with much higher precision.
Of particular interest will be  measurements of differential distributions.

\Acknowledgements
I would like to thank the CKM2010 organiser for partial financial support.

\end{document}